\documentclass[10pt,journal,compsoc]{IEEEtran}

\usepackage{booktabs}
\usepackage{subcaption}
\usepackage{listings}
\usepackage{threeparttable}
\usepackage{multirow}
\usepackage{mwe,tikz}
\usepackage{textcomp}
\usepackage{dblfloatfix}    
\usepackage{siunitx}
\usepackage{xcolor,colortbl}
\usepackage{xparse}
\usepackage{enumitem}
\usepackage{url}

\ifCLASSOPTIONcompsoc
  \usepackage[nocompress]{cite}
\else
  \usepackage{cite}
\fi

\newcommand{\titlename}{$f$\kern-0.18em\emph{unc}\kern-0.05em X}
\newcommand{\name}{$f$\kern-0.18em\emph{unc}\kern-0.05em X}

\newcommand{\parsl}{Parsl}

 \definecolor{deepred}{rgb}{0.6,0,0}
 \definecolor{deepgreen}{rgb}{0,0.5,0}

\lstdefinestyle{PythonStyle}
{
        language=Python,
        frame=tb,
        basicstyle=\ttfamily\footnotesize,
        upquote=true,
        numbers = none,
        escapechar=`,
        float,
        moredelim=[il][]{--latexlabel},
        otherkeywords={self, def},             
        commentstyle=\color{blue},
        keywordstyle=\bfseries\color{black},
        emph={FuncXClient,register_function,run, automo_preview,get_result,GlobusFile,get_redis_client, extract_metadata, solve, process_stills}, 
        emphstyle=\bfseries\color{deepred},    
        showstringspaces=false            %
}

\lstdefinestyle{PythonStyleInLine}
{
        language=Python,
        basicstyle=\small\ttfamily,
        upquote=true,
        numbers = none,
        numberstyle=\footnotesize,
        escapechar=`,
        moredelim=[il][]{--latexlabel},
        otherkeywords={self},             
        commentstyle=\color{blue},
        keywordstyle=\bfseries\color{red},
        emph={MyClass,__init__,@python_app,@bash_app},          
        emphstyle=\bfseries\color{deepred},    
        showstringspaces=false,            %
        frame=bt
}

\newif\iffinal

\finaltrue

\iffinal
  \newcommand{\ian}[1]{}
  \newcommand{\ryan}[1]{}
  \newcommand{\kyle}[1]{}
  \newcommand{\zhuozhao}[1]{}
  \newcommand{\yadu}[1]{}
  \newcommand{\tyler}[1]{}
  \newcommand{\katznote}[1]{}
  \newcommand{\ben}[1]{}
  \newcommand{\anna}[1]{}
  \newcommand{\add}[1]{}
\else
  \newcommand{\ian}[1]{{\textcolor{red}{ Ian: #1 }}}
  \newcommand{\ryan}[1]{{\textcolor{magenta}{ Ryan: #1 }}}
  \newcommand{\kyle}[1]{{\textcolor{purple}{ Kyle: #1 }}}
  \newcommand{\zhuozhao}[1]{{\textcolor{cyan}{ Zhuozhao: #1 }}}
  \definecolor{darkgreen}{rgb}{0,0.5,0}
  \newcommand{\tyler}[1]{{\textcolor{green}{ Tyler: #1 }}}
  \newcommand{\katznote}[1]{{\textcolor{darkgreen}{ Dan: #1 }}}
  \newcommand{\yadu}[1]{{\textcolor{orange}{ Yadu: #1 }}}
  \definecolor{pink}{rgb}{1.0,0,0.5}
  \newcommand{\ben}[1]{{\textcolor{deepgreen}{ Ben: #1 }}}
  \newcommand{\anna}[1]{{\textcolor{deepred}{ Anna: #1 }}}

\fi

\newif\ifchange

\ifchange
  
\else
  
\fi

\begin{document}

\title{\name{}: Federated Function as a Service for Science}

\author{Zhuozhao~Li, Ryan~Chard, Yadu~Babuji, Ben~Galewsky, Tyler~Skluzacek, Kirill~Nagaitsev, Anna~Woodard, Ben~Blaiszik, Josh~Bryan, Daniel~S.~Katz,~\IEEEmembership{Senior Member,~IEEE,} Ian~Foster,~\IEEEmembership{Fellow,~IEEE,} and~Kyle~Chard
\IEEEcompsocitemizethanks{
\IEEEcompsocthanksitem Z. Li is with Department of Computer Science and Engineering and Research Institute of Trustworthy Autonomous Systems, Southern University of Science and Technology, Shenzhen, China. Email: lizz@sustech.edu.cn
\IEEEcompsocthanksitem Y. Babuji, T. J. Skluzacek, A. Woodard, B. Blaiszik, J. Bryan, I. Foster, and K. Chard are with the University of Chicago, Chicago, IL, 60637. E-mail: \{yadunand, skluzacek, annawoodard, blaiszik, jbryan, foster, chard\}@uchicago.edu
\IEEEcompsocthanksitem R. Chard and I. Foster are with Argonne National Laboratory, Lemont, IL 60439. E-mail: \{rchard, foster\}@anl.gov
\IEEEcompsocthanksitem D. S. Katz is with NCSA, CS, ECE, and the iSchool, University of Illinois, Urbana, IL, 61801. E-mail: d.katz@ieee.org
\IEEEcompsocthanksitem B. Galewsky is with NCSA, University of Illinois, Urbana, IL, 61801. E-mail: bengal1@illinois.edu
\IEEEcompsocthanksitem K. Nagaitsev is with Northwestern University, Evanston, IL, 60208. E-mail: knagaitsev@u.northwestern.edu
}}

%
%

\IEEEtitleabstractindextext{%
\begin{abstract}
\name{} is a distributed function as a service (FaaS) platform that enables flexible, 
scalable, and high performance remote function execution. 
Unlike centralized FaaS systems, \name{} decouples the cloud-hosted management functionality from the edge-hosted execution functionality. 
\name{}'s endpoint software can be deployed, by users or administrators, on arbitrary laptops, clouds, clusters, and supercomputers,
in effect turning them into function serving systems. 
\name{}'s cloud-hosted service provides a single location for registering, sharing, 
and managing both functions and endpoints. It allows for transparent, secure, 
and reliable function execution across the federated ecosystem of endpoints---enabling
users to route functions to endpoints based on specific needs. 
\name{} uses containers (e.g., Docker, Singularity, and Shifter) to provide common
execution environments across endpoints. 
\name{} implements various container management strategies to execute functions with high performance and efficiency on diverse \name{} endpoints. 
\name{} also integrates with an in-memory data store and Globus for managing data that may span endpoints. %
We motivate the need for \name{}, present 
our prototype design and implementation, 
and demonstrate, via experiments on two supercomputers,
that \name{} can scale to more than \num{130000} concurrent workers. We show that \name{}'s container warming-aware routing algorithm can reduce the completion time for 3000 functions by up to 61\% compared to a randomized algorithm and the in-memory data store can speed up data transfers by up to 3x compared to a shared file system.
\end{abstract}

}

\maketitle

\IEEEdisplaynontitleabstractindextext

%
\IEEEpeerreviewmaketitle

\IEEEraisesectionheading{\section{Introduction}}

\IEEEPARstart{T}{he} exponential growth of data and increasing hardware diversity is driving the need for computation to occur wherever it makes the most sense, for example, on a suitable computer, where particular software is available, or near data. 
Prior research, in grid~\cite{Foster2001} and peer-to-peer~\cite{milojicic2002peer} computing, 
has studied and explored the foundations for remote computing. 
However, with the exception of cloud platforms, general-purpose remote computation has remained elusive due to, for example, slow and unreliable network communications, security challenges, and dependencies between software and heterogeneous computer architectures.

Commercial cloud providers have been at the forefront of 
recent advances in networks, hardware, and distributed computing, leveraging
widespread virtualization, universal trust fabrics, 
and high-speed networks to deliver serverless computing
services such as function-as-a-service (FaaS)~\cite{baldini2017serverless,fox2017status,varhese19cloud}. 
FaaS enables developers to register a high-level programming 
function and to then invoke that function many times
by passing input arguments. The user needs not concern
themselves with provisioning infrastructure or 
configuring execution environments. 
FaaS systems have quickly become integral to a wide
range of applications, particularly for event-based
and dev-ops applications. 

The FaaS model is particularly attractive in science
as a way of decomposing monolithic science applications into
a collection of modular, performant, and extensible functions~\cite{foster2017cloud,spillner2017faaster,malawski2016towards,fox2017conceptualizing,kiar2019serverless}. 
However, existing FaaS systems are typically centralized and specific to a particular cloud, rather than being designed to be deployed
on heterogeneous research cyberinfrastructure (CI) or to use federated resources.
Typically research CI uses batch scheduling interfaces and inflexible authentication and authorization models, 
which does not lend itself to the fine-grain and sporadic function workloads. In response, we propose a federated FaaS model for general-purpose remote computing at scale across diverse CIs, both centrally and at the edge. 

In this paper, we present \name{}, a federated, scalable, 
and high-performance function execution platform. 
\name{} leverages a \emph{distributed endpoint model} to support remote function execution across distributed and heterogeneous research CI. 
Users can transform many computing resources, such as laptops, clouds, clusters, supercomputers, or Raspberry Pis they are authorized to access, into function serving systems by deploying \name{}'s endpoint software. 
Users then use the cloud-hosted \name{} service to register Python functions and invoke those functions on their deployed endpoints. 
\name{} manages the reliable and secure execution of functions, staging function code and inputs, provisioning resources, managing safe and secure
execution (optionally in containers), monitoring execution, and returning outputs to users.  Thus, users benefit from the convenience
and reliability of a cloud-hosted service combined with the 
flexibility and performance of a federated ecosystem of endpoints.

We extend our previous work~\cite{chard2020funcx} 
to support complex data dependencies between scientific functions.
Specifically, we focus on enabling 
data transfer between functions that are executing
on the same (\textit{intra-endpoint}) or different
(\textit{inter-endpoint}) endpoints. 
For intra-endpoint communication we use an in-memory data store, for inter-endpoint communication we use Globus~\cite{foster2011globus}.
We also present new heuristic-based container management and function routing schemes that reduce container warming overhead and 
efficiently route functions to appropriately configured containers.

The primary novelty of our work is in the adaptation of the FaaS paradigm to a federated research ecosystem, combining a distributed endpoint model with a hosted FaaS platform to support remote function execution across distributed and heterogeneous research CI. 
We demonstrate the viability of our approach with a highly modularized and extensible design as well as a  scalable and performant implementation. We also show that it is beneficial to decompose scientific applications into monolithic functions that may be executed on different remote resources. The contributions of our work are as follows: 

\begin{itemize}
    \item \name{}, a distributed and federated FaaS platform that can: 
          be deployed on research CI, dynamically provision and manage
					resources, leverage various container technologies, and facilitate 
					secure, scalable, and distributed function execution.
    \item Automated data movement between functions using widely-used in-memory data stores and high-performance data transfer technology to transparently support data dependencies between functions.
    \item Design and evaluation of performance enhancements for function
          serving on distributed research CI, including function warming, 
					batching, and function routing.
    \item Experimental studies showing that \name{} delivers execution 
					latencies comparable to those of commercial FaaS platforms 
          and scales to 1M+ functions across 130K active workers on supercomputers.
\end{itemize}

The rest of this paper is as follows. 
\S\ref{sec:requirements} describes an example use case and presents general requirements for FaaS in science.
\S\ref{sec:funcx} presents a conceptual model of \name{}.
\S\ref{sec:arch} describes the \name{} system architecture.
\S\ref{sec:data_management} discusses how data is managed in \name{}.
\S\ref{sec:routing} presents \name{}'s container management model.
\S\ref{sec:evaluation} evaluates \name{} performance. 
\S\ref{sec:discussion} reviews \name{}'s use in scientific case studies. 
\S\ref{sec:survey} discusses related work.
Finally, \S\ref{sec:conclusion} summarizes our contributions.

\section{Motivations and Requirements} \label{sec:requirements}

Over the last two years the scientific community has been working to understand 
SARS-CoV-2 and develop effective tests, therapeutics, and vaccines.
However, progress in these areas is dependent on our ability
to understand SARS-CoV-2 protein structures. 
At Argonne's Advanced Photon Source~\cite{kim2020crystal}, scientists use an emerging method called fixed-target serial synchrotron crystallography (SSX)  to collect physiological temperature data from thousands of protein crystals.

Data are generated at unprecedented rates with tens of thousands of images captured each hour. Keeping pace with the experiment requires rapid data processing across multiple, heterogeneous 
computing resources to efficiently analyze, refine, solve, and 
curate structures.

To meet these 
data processing and publication needs, SSX scientists have adopted an automated
data management framework~\cite{Wilamowski2020} that can manage data acquisition, 
analysis, curation, and visualization. 
Throughout this workflow, there are needs for computation both
at the edge to detect and pre-process data rapidly, as well as on 
HPC resources to perform computationally expensive analysis tasks and produce structures.
\emph{Each of these steps relies on different packages and functions, has different processing 
durations, occurs at different times, and requires different types and amounts of computing resources.
Thus, it is essential that the scientists be able to decompose the entire processing pipeline 
into a series of individual functions to perform on data as they are moved and transformed.}
These functions, shown in Listing~\ref{lst:ssx}, analyze 
individual images, refine and solve the crystal structure, 
and extract metadata and create plots before publishing results.

\begin{lstlisting}[style=PythonStyle, caption={Three functions used in the SSX pipeline and an example of how the funcX SDK is used to register and invoke the \texttt{process\_stills} function. \label{lst:ssx}}]
def process_stills(data):
  inputs = data['inputs']
  phil = data['phil']
  cmd = f'dials.stills_process {phil} {inputs}'
  res = subprocess.run(cmd)
  return res.stdout
  
def solve(data):
  from gladier.tools import template_prime
  pdata = template_prime.substitute(data['template'])
  cmd = f"prime.run {pdata} > prime.log"
  res = subprocess.run(cmd)
  return res.stdout
  
def extract_metadata(data):
  from gladier.tools import (get_dims, 
    get_lattice_counts, plot_lattice_counts)
  lc = get_lattice_counts(xdim, ydim, int_files)
  plot_name = f'1int-sinc.png'
  plot_lattice_counts(xdim, ydim, lc, plot_name)
  return plot_name
  
fc = FuncXClient()
func_id = fc.register_function(process_stills)
endpoint_id = '863d-...-d820d'

input_data = {'inputs': '...', 'phil': '...'}
task_id = fc.run(func_id, endpoint_id, 
                 data=input_data)
res = fc.get_result(task_id)
  
\end{lstlisting}

This typical science use case, with parallels in many
other domains described in our previous work~\cite{chard2020funcx}, 
highlights the benefits of 
FaaS approaches (e.g., decomposition, abstraction, 
flexibility, scalability, reliability), 
and also elucidates several requirements for FaaS approaches.

\begin{itemize} 

     \item \textbf{Research CI}: 
     functions may require HPC-scale and/or specialized and heterogeneous resources. Many resources expose batch scheduler interfaces (with long delays, periodic downtimes, proprietary interfaces) and specialized container technology (e.g., Singularity, Shifter) that make it challenging to provide common execution interfaces, on-demand and elastic capacity, and fault tolerance.
     \item \textbf{Distribution}: different parts of an application may be most efficiently executed on different, often distributed, resources (e.g., near data, on a specialized computer).
	\item \textbf{Data}: functions analyze both small and large data, stored in various locations and formats, and accessible via different methods (e.g., Globus~\cite{chard14efficient}).
    \item \textbf{Authentication}: institutional identities and specialized security models are used to access data, compute resources, and other cyberinfrastructure.
    \item \textbf{State}: functions may be connected and share state (e.g., files or database connections) to decrease overheads. 
\end{itemize}

Existing FaaS solutions may satisfy these requirements partially, but not completely. 
For example, some FaaS systems (e.g., OpenWhisk~\cite{openwhisk}, KNIX~\cite{knix}) support on-premise deployments on specialized hardware (e.g, GPU), but not on distributed and federated computing resources.
Some FaaS systems (e.g., DFaaS~\cite{ciavotta2021dfaas}, ChainFaaS~\cite{ghaemi2020chainfaas}) support function execution in distributed environments, but not on research CI. 
Here we present \name{}, a federated and scalable FaaS platform that enables researchers to decompose applications into functions and execute them on arbitrary remote computers via the FaaS paradigm.
\section{Conceptual Model}\label{sec:funcx}

We first describe the conceptual model behind \name{}
to provide context to the implementation architecture. 
\name{} allows users to register and then execute \emph{functions} 
on arbitrary \emph{endpoints}.
All user interactions with \name{} are performed via a 
REST API implemented by a cloud-hosted \name{} service. 

\textbf{Functions:}
\name{} is designed to execute \emph{functions}: snippets of Python code
that perform some activity. A \name{} function explicitly defines
a Python function and input signature. 
The function body must specify all imported modules. While \name{} supports only Python functions, users can easily write Python functions to invoke tools written in other languages.
Listing~\ref{lst:ssx} shows several functions used in the SSX pipeline mentioned in \S\ref{sec:requirements}. The \texttt{process\_stills} function takes a single input dictionary as input, which includes the locations of the images and the
\textit{phil} file describing the analysis configuration. The function then makes use of the \textit{DIALS}~\cite{winter2018dials} tool to analyze the image.

\textbf{Function registration:}
A function must be registered with \name{} before it can be executed. 
The registration includes a name and the serialized function body. 
Optionally, it 
may also specify users, or groups of users, who may be authorized to invoke the function, and a container image to be used for execution. Containers 
allow the construction of 
environments with the 
dependencies
(system packages and Python libraries) required to execute the function.
\name{} assigns a universally
unique identifier (UUID) for management and invocation. Users may update
functions they own. 

\textbf{Endpoints:}
A \name{} endpoint is a logical entity that represents a compute resource.
The corresponding \name{} agent allows the \name{} service to dispatch 
functions to that resource for execution. The agent handles 
authentication and authorization, provisioning of nodes on the
compute resource, and monitoring and management. Administrators
or users can deploy a \name{} agent and register an endpoint for themselves and others,
providing descriptive (e.g., name, description) metadata. 
Each endpoint is assigned a UUID for subsequent use.

\textbf{Function execution:}
Authorized users may invoke a registered function on a selected endpoint. 
To do so, they issue a request via \name{} that 
identifies 
the function and endpoint to be used as well as inputs
to be passed to the function.   
Functions are executed asynchronously: each invocation returns an
identifier via which progress may be monitored and results retrieved. 
\emph{In this paper, we refer to an invocation of a function as a ``task.''} Importantly, following the FaaS model, while users must specify the specific endpoint for use, they do not manage the resources on which the function is executed (e.g., nodes, containers, or  modules)

\textbf{\name{} service:}
Users interact with \name{} via a cloud-hosted service that
exposes a REST API for registering functions and 
endpoints, and for executing functions, monitoring their execution, and retrieving results. 
The REST API provides a uniform interface via which users can
make asynchronous and stateless calls to manage endpoints and function executions. REST APIs are the the most common interface for FaaS platforms (e.g., AWS Lambda~\cite{amazonlambda} and Google Cloud Functions~\cite{googlecloudfunctions}).
The service is connected to 
accessible endpoints via the endpoint
registration process.

\textbf{User interface:}
\name{} provides a Python SDK that wraps the REST API. 
Listing~\ref{lst:ssx} shows an example of how the SDK can be used
to register and invoke a function on a specific endpoint.
The example first constructs a \emph{client} and registers the \texttt{process\_stills} function. 
It then invokes the registered function using the \texttt{run}
command, passing the unique function identifier, 
the endpoint id on which to execute the function, and inputs
(in this case \texttt{data}). 
Finally, the example shows that the results can be asynchronously retrieved
using \texttt{get\_result}.

\section{Architecture and Implementation}\label{sec:arch}
\begin{figure}[h]
\vspace{-0in}
  \includegraphics[width=\columnwidth,trim=0.2in 0.1in 1.6in 0in,clip]{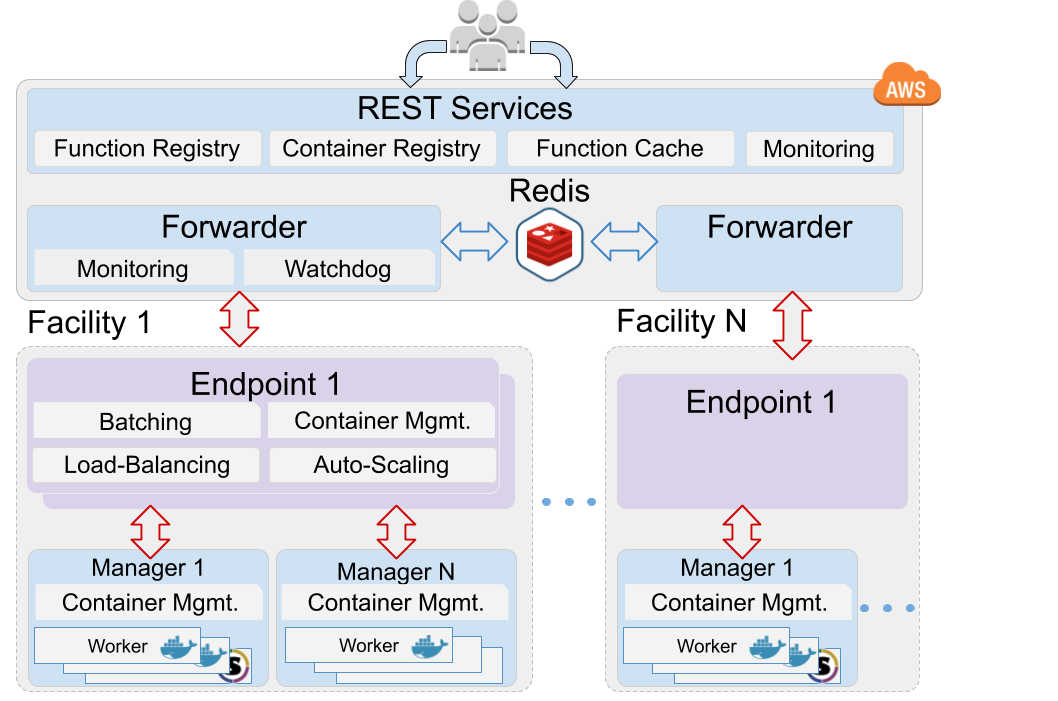}
  \caption{\name{} architecture, showing the \name{} service (top) consisting of a REST interface, Redis store, and Forwarders. \name{} endpoints (bottom) provision resources and coordinate the execution of functions. 
  }
\label{fig:arch}
\vspace{-0in}
\end{figure}

\name{} combines a cloud-hosted management service with software agents
deployed on remote resources: see~\figurename{~\ref{fig:arch}}.

\subsection{The \name{} Service}
The \name{} service maintains a registry of \name{} endpoints,
functions, and users in a persistent AWS Relational Database Service (RDS) database. 
To facilitate rapid function dispatch, \name{} stores serialized function codes 
and tasks (including inputs and task metadata) in an AWS ElastiCache Redis hashset.
The service also manages a Redis queue for each endpoint that stores task IDs for
tasks to be dispatched to that endpoint.
The service exposes a REST API to register and manage endpoints, 
register functions, execute and monitor functions, and retrieve the output from tasks. 
The \name{} service is secured using Globus Auth~\cite{GlobusAuth}, 
which allows users, programs and applications, and \name{} endpoints
to securely make API calls. 
When an endpoint registers with the \name{} service,
a unique \emph{forwarder} process is created for each 
endpoint. 
Endpoints establish secured ZeroMQ connections with their 
forwarder to receive tasks, return results, and perform heartbeats.

\name{} implements a hierarchical task queuing architecture consisting 
of queues at the \name{} service, endpoint, and worker. These queues support 
reliable fire-and-forget function execution that is resilient to failure 
and intermittent endpoint connectivity.  
At the first level, 
each registered endpoint is allocated a unique Redis \emph{task queue} 
and \emph{result queue} to store and track tasks, which are implemented using Redis \texttt{Lists} structure.
We use Redis as it provides a simple yet performant system for brokering tasks. Redis is offered as a hosted Amazon service and can be elastically scaled as workload increases. \name{} serves primarily as a broker to manage and distribute tasks. Redis provides high throughput queuing via an in-memory store with little overhead on the tasks and results passed through the queue---an important requirement for providing low latency execution. One limitation of this approach is that we must implement message acknowledgments to ensure that tasks and results are communicated reliably between clients, endpoints, and the \name{} service. We note that as use cases expand, we may need to consider other message queues, such as Kafka~\cite{kafka}, Pulsar~\cite{pulsar}, or AMQP-based systems (e.g., RabbitMQ~\cite{rabbitmq}).

\begin{figure}[h]
  \centering
  \includegraphics[width=0.88\columnwidth]{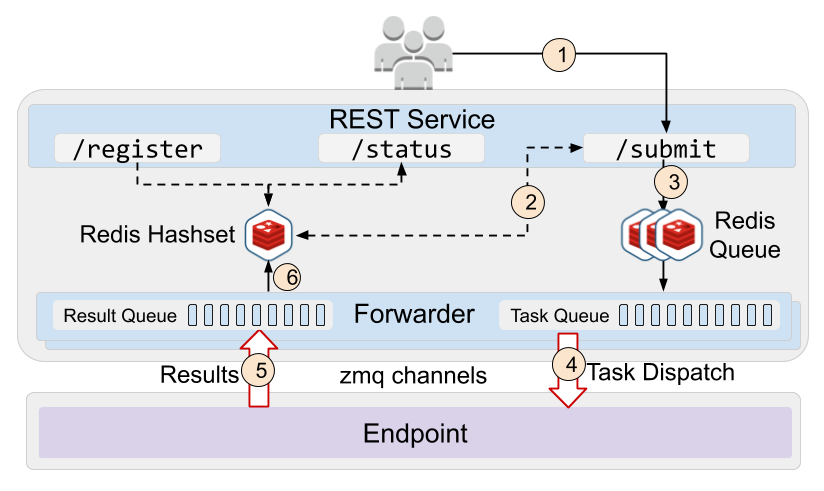}
  \caption{\name{} task execution path. A task submitted to \name{} (1) is stored in Redis (2), queued for execution (3), and dispatched via a Forwarder to an endpoint (4); results are returned (5), then stored in Redis for users to retrieve (6). 
  }
\label{fig:taskpath}
\end{figure}

\figurename~\ref{fig:taskpath} shows the \name{} task lifecycle.
At function submission, the \name{} service routes the task to the specified endpoint's task queue.
The forwarder listens to the queue for tasks and then 
dispatches the task to the corresponding agent.
\name{} agents internally queue tasks at both the manager and worker.
These queues ensure that tasks are not lost once they
have been delivered to the endpoint. 
Results are returned to the \name{} service and stored in the endpoint's
result queue until they are retrieved by the user.

\name{} relies on AWS-hosted databases, caches, and Web serving infrastructure to reduce operational overhead,
elastically scale resources, and provide high availability. 
While these services provide significant benefits to \name{}, they have associated costs. 
To minimize these costs we apply several techniques, such as
using small cloud instances with responsive 
scaling to minimize the steady state
cost 
and restricting the size of input and output data 
passed through the 
\name{} service to reduce storage (e.g., in Redis store) and data egress costs. 
Further, we periodically purge results from the Redis store 
once they have been retrieved by the client or after a period 
of time. 

The \name{} service is designed to provide robustness and fault tolerance using several techniques. First, the \name{} service implements health checks, including a liveness check, CPU utilization, and response time monitoring, etc. The service is automatically restarted when these health checks indicate failures. Second, the RDS database and Redis task queue are replicated to ensure that any data (e.g., users, functions, endpoints, and tasks) are not lost. Third, the forwarder uses configurable periodic heartbeats (30 seconds by default) to detect if an agent is disconnected. A task is sent to an agent only when it is connected. When an agent is disconnected, all the tasks dispatched to the agent are returned back into the task queue and the tasks are forwarded to that agent when it reconnects. Fourth, tasks are cached at each layer and only removed when downstream layers have acknowledged receipt.
Finally, to serve the distributed endpoints in \name{}, the \name{} service is deployed on cloud-hosted services, which internally provide high reliability and robustness. 

\subsection{Function Containers} 
\name{} uses containers to package function code and dependencies that are to be deployed
on a compute resource. 
Our review of container technologies, including Docker~\cite{merkel2014docker}, LXC~\cite{LXC}, Singularity~\cite{kurtzer2017singularity}, Shifter~\cite{jacobsen2015contain}, and CharlieCloud~\cite{priedhorsky2017charliecloud}, 
led us to adopt Docker, Singularity, and Shifter. 
Docker works well for local and cloud deployments, whereas
Singularity and Shifter are designed for use in HPC environments
and are supported at large-scale computing facilities (e.g., Singularity at ALCF and Shifter at NERSC). 
Singularity and Shifter implement similar models and thus 
it is easy to convert from a common representation (e.g., a Dockerfile)
to both formats. 

\name{} requires
that each container includes a base set of software, including Python~3 
and \name{} worker software.  Other system libraries or Python 
modules needed for function execution must also be included. 
When registering a function, users may optionally specify a container to be used for execution;
if no container is specified, \name{} executes functions using the worker's Python environment. 
In future work, we intend to make this process dynamic, using repo2docker~\cite{repo2docker}
to build Docker images and convert them to site-specific container formats
as needed.

\subsection{The \name{} Endpoint}\label{sec:executor}

The \name{} endpoint represents a remote resource and 
delivers high-performance remote execution of 
functions in a secure, scalable, and reliable manner.

The endpoint architecture, depicted in the lower portion of \figurename{~\ref{fig:arch}}, 
is comprised of three components, which are discussed below:
\begin{itemize}
\item \emph{\name{} agent}: a persistent process that queues and forwards tasks and results, 
interacts with resource schedulers, and load balances tasks.
\item \emph{Manager}: manages the resources for a single compute node on an endpoint by deploying and managing a set of workers.
\item \emph{Worker}: executes tasks (optionally within a container).
\end{itemize}
The \emph{\name{} agent} is a software agent that is deployed by a user on a 
compute resource (e.g., an HPC login node, cloud instance, or a laptop).
It registers with the \name{} service and acts 
as a conduit for routing tasks and results between the service and workers. 
The \name{} agent manages resources on its system
by working with the local scheduler or cloud API to deploy 
\emph{managers} on compute nodes.
The \name{} agent uses a pilot job model~\cite{turilli2018comprehensive} to provision 
and communicate with resources in a uniform manner,
irrespective of the resource type (cloud or cluster) or local resource manager (e.g., Slurm, PBS, Cobalt).
As each manager is launched on a compute node, it connects to and registers with the \name{} agent. 
The \name{} agent then uses ZeroMQ sockets to communicate 
with its managers. 
To minimize blocking, all communication is performed by threads
using asynchronous communication patterns. 
The \name{} agent uses a randomized scheduling algorithm to allocate tasks 
to suitable managers with available capacity.
The \name{} agent can be configured to provide 
access to specialized hardware or accelerators. 
When deploying the agent, users can specify how worker
containers should be launched, enabling them to mount
specialized hardware and execute functions on that hardware. 
In future work, we will
extend the agent configuration to specify custom hardware and 
software capabilities and report
this information to the \name{} agent and service for scheduling.

To provide fault tolerance and robustness, for example with respect to 
node failures, the \name{} agent relies on periodic heartbeat messages and a 
process to detect lost managers. The \name{} agent tracks 
tasks that have been distributed to managers so that when failures
do occur, lost tasks can be re-executed (if permitted). 
\name{} agents communicate with the \name{} service's forwarder
via a ZeroMQ channel. 
Loss of a \name{} agent is detected by the forwarder and when 
the \name{} agent recovers, it repeats the registration 
process to acquire a new forwarder and continue receiving tasks.
To reduce overheads, the \name{} agent can shut down managers to release resources when they are not
needed, suspend managers to prevent further tasks from being scheduled to them,
and monitor resource capacity to aid scaling decisions.

\emph{Managers} represent, and communicate on behalf of, the 
collective capacity of the workers on \emph{a single node}, using just two sockets per node. 
Managers determine 
the available CPU and memory resources on a node, and partition the
node among the workers. 
Managers advertise deployed container 
types and available capacity to the endpoint. 

\emph{Workers} persist on a node (optionally within containers) and each executes one task 
at a time. Since workers have a single responsibility,
they use blocking communication to wait for tasks from the manager. 
Once a task is received, it is deserialized, executed,
and the serialized results are returned via the manager.

\subsection{Managing Compute Infrastructure} \label{sec:provider}
\name{} is designed to support a range of computational resources, from 
embedded computers to clusters, clouds, and supercomputers,
each with distinct access modes.
As \name{} workloads are often sporadic, resources must be provisioned and deprovisioned
as needed to reduce costs due to idle resources.
\name{} uses \parsl{}'s provider interface~\cite{babuji19parsl} 
to interact with various resources, specify resource-specific
requirements (e.g., allocations, queues, limits, cloud instance types), 
and define rules for automatic scaling (i.e., limits and scaling aggressiveness).
This interface allows \name{} to be deployed on batch schedulers 
such as Slurm, PBS, Cobalt, SGE, and Condor; 
major cloud systems 
such as AWS, Azure, and Google Cloud; 
and Kubernetes.

\subsection{Serialization}
\label{sec:serialization}

\name{} supports the registration of arbitrary Python functions and the 
passing of data (e.g., primitive types and complex objects) to/from those functions. 
\name{} uses a Facade interface with
several serialization libraries (including pickle, dill, and JSON) as some Python object types cannot be serialized with some serialization libraries, and no single serialization library can serialize all objects.
The \name{} serializer sorts the serialization libraries by speed and applies
them in order successively until the object is successfully serialized. 
This approach combines the strengths of various libraries, including
support for complex objects (e.g., machine learning models) and traceback objects in a fast and transparent fashion.
Once objects are serialized, they are packed into buffers with headers that include routing tags and the serialization method,
such that only the buffers need to be unpacked and deserialized at the destination.

\subsection{Batching}
\name{} supports two batching to amortize costs 
across many function requests:
internal batching enables managers to request many tasks on behalf of their workers, minimizing network communication costs; 
and, user-facing \emph{batching} that
enables users to define batches of function inputs, 
allowing users to trade off 
efficient execution and increased 
per-function latency by creating fewer, larger
requests. The SDK includes a matching batch
interface for retrieving the results of many
tasks concurrently.

\subsection{Security Model}
\name{} requires a security model to ensure that functions are executed on endpoints by authenticated and authorized users and that one function cannot interfere with another. 

\textbf{Authentication and authorization:}
Since \name{} endpoints may be deployed across arbitrary resources, we first summarize authentication and authorization requirements.

\begin{itemize}
    \item Different research CI may rely on diverse identity management systems and authentication models (e.g., two-factor authentication). To ease the deployment of \name{} agent on any resources, \name{} needs a general model that provides a uniform API, rather than maintaining a set of APIs for the diverse identity providers.
    \item Users may have to use different accounts (e.g., institution accounts, national CI credentials, or national laboratory accounts) to access different resources. Users would like to use one account to authenticate \name{} endpoints infrequently.
    \item One frequent use case in scientific computing is that resources are shared among a group of scientists. Ideally, the authorization model should enable users to grant access to others while enforcing secure delegation.
\end{itemize}

\name{} uses Globus Auth~\cite{GlobusAuth} for identity and access management (IAM), and protection of all APIs. 
We use Globus Auth as it satisfies the above requirements, is widely adopted in scientific community, implements standard protocols (e.g., OAuth~2), enables simple delegation (e.g., such that a user may allow the \name{} service or another user to access their endpoint), 
and offers a flexible OAuth client model 
for developing the \name{} SDK.
Although Globus Auth is used as the primary implementation, other IAM services that provide similar capabilities and interfaces could be integrated with \name{}.

The \name{} service is registered as a Globus Auth \emph{resource server}, allowing users to 
authenticate using a supported identity (e.g., institution, Google, ORCID)
and enabling various OAuth-based authentication flows (e.g., native client)
for different scenarios. 
\name{} has associated Globus Auth scopes
(e.g., ``urn:globus:auth:scope:funcx:register\_function'')
via which other clients (e.g., applications and services) 
may obtain authorizations for programmatic access. 
\name{} endpoints are themselves Globus Auth native clients, each dependent on
the \name{} scopes, which are used to securely connect to the \name{} service. 
Endpoints require the administrator to authenticate prior to registration
in order to acquire access tokens used for constructing API requests.
The connection between the \name{}
service and endpoints is established using ZeroMQ. Communication addresses
are sent 
as part of the registration process. Inbound
traffic from endpoints to the cloud-hosted service is limited to known IP addresses. 

\textbf{Isolation:}
\name{} function execution can be isolated in containers to ensure that functions 
cannot access data or devices outside their context.
To enable fine-grained tracking of execution, 
we store execution request histories in the \name{} service
and in logs on \name{} endpoints.

\section{Data Management}\label{sec:data_management}
Data management is essential for many 
applications: 
functions may interact with large and/or remote datasets, and tasks may use the outputs of other tasks as inputs.
This section describes how data can be staged and managed between different \name{} endpoints (\emph{inter-endpoint}) and between different functions within an endpoint (\emph{intra-endpoint}).

\subsection{Inter-endpoint Data Transfers}
To minimize operational costs and performance overheads we limit the size of 
data that can be passed through the \name{} service to 10~MB. To enable functions to be seamlessly invoked with large data that may be located on remote
computers, we require an out-of-band data transfer mechanism. 
We summarize the primary requirements as follows. 

\begin{itemize}
    \item Transfers can be managed programmatically by \name{}.
    \item The transfer mechanism should be natively supported and approved by the administrations of research CI.
    \item Transfers should be optimized and provide high performance, endpoint-to-endpoint movement.
    \item The transfer mechanism should be interoperable with \name{}'s authentication and authorization model (i.e., Globus Auth) 
    to secure data transfers on behalf of users. 
    \item The transfer mechanism should allow a user's functions to fetch data that is shared among a group.
\end{itemize}

\begin{lstlisting}[style=PythonStyle, caption={Inter-endpoint data transfer with Globus. \label{lst:globus}}]
from funcx import GlobusFile

data = GlobusFile(globus_endpoint_id='65e...', 
                  file_path='/~/file.txt')
task_id = fc.run(func_id, endpoint_id, 
                 remote_data=data)
\end{lstlisting} 

We focus on wide area data management, rather than cloud storage, as data may be stored or generated in different locations (e.g., instruments, campus clusters, supercomputers) in many scientific use cases. 
Based on the requirements above, we integrate Globus transfer~\cite{chard14efficient} to streamline inter-endpoint data transfers. 
Globus has several advantages that lead us to this choice: i) it is a research data management service that provides high-performance data transfers between arbitrary storage resources, such as supercomputers, laptops, and clouds; ii) it is widely deployed on research CI and used in scientific research; iii) data are transferred directly between the source and destination systems via the GridFTP~\cite{allcock2005globus} protocol;
iv) it provides a Python SDK that allows a user's functions to fetch shared data.

To use Globus, the Globus Connect software 
must be installed on the storage system, this is often done by administrators installing Globus
Connect Server on large clusters or it can be done individually in user-space using Globus Connect Personal. 
Storage systems are registered as a Globus endpoint with associated authentication mechanism
in the Globus service. Each endpoint is given a unique endpoint identifier that is used when transferring data.

In this paper, we extend \name{} to allow for references to Globus-accessible files to be passed as input/output
to/from a function. Specifically, users must specify the Globus endpoint and the path
to the file on that endpoint. 
When Globus-accessible files are passed 
to/from a \name{} function,
\name{} can automatically stage data either prior to, or after invocation of the function.
An example 
of using Globus for inter-endpoint data transfer is shown in Listing~\ref{lst:globus}.

We have found that Globus is well suited for our current use cases; however, other mechanisms (e.g., HTTP, FTP, and rsync) could also be used for inter-endpoint data transfers by augmenting functions to make direct data downloads or uploads. In future work we will extend the inter-endpoint transfer model in \name{} to transparently support these mechanisms as we have done in Parsl.

\subsection{Intra-endpoint Data Transfers}\label{sec:intra}
Modern applications may involve frequent fine-grained communications 
among functions executed on an individual endpoint (i.e., intra-endpoint data transfers). 
For example, distributed machine learning (ML) training may require that state be coordinated 
among all worker nodes; and MapReduce-style applications often 
involve a shuffle phase where every map task sends data to every reduce task. 

Here we describe the advantages and disadvantages of potential intra-endpoint data management approaches.

\begin{itemize}
\item \textbf{A shared file system} that can be accessed by every worker on an endpoint. The effort to attach such storage to a \name{} endpoint is minimal, as many clusters, clouds, and supercomputers provide built-in shared file system (sharedFS) or object storage. However, they often have high access cost, limited IO performance, and high latency when writing and reading many files. 
\item \textbf{MPI} is a message passing fabric that is  highly scalable and optimized for data communications on supercomputers with specialized interconnects;
however, MPI libraries are not natively available or optimized for many computers (e.g., clouds and private clusters). More importantly, the synchronous nature of MPI's collective communication is not well-suited for the asynchronous task-based model in \name{}, as it blocks tasks from making progress even when partial results are ready, which is important for many performance-driven asynchronous applications (e.g., distributed machine learning training); HPC containers often must be adapted to make use of local MPI libraries; and a failure of one MPI process may cause other MPI processes to block, which stops other tasks from continuing. We note that fault tolerance has improved in the recent release of MPI 4.0; however, this is not commonly deployed at the time of writing.
\item \textbf{Socket and socket-like connections} (e.g., ZeroMQ) between workers can
provide high throughput and low latency direct data transfers. However, creating pair-wise connections
between workers is expensive and in some cases workers (e.g., in containers) may not be network addressable or may not have sufficient open ports to support connections between all workers.
\item \textbf{In-memory data stores} (e.g., MemCache~\cite{nishtala2013scaling} and Redis~\cite{redis}) provide
higher IOPS and lower latency than shared file systems and support more data types than socket connections (e.g., serialized data). However, they require that storage be provisioned explicitly, that additional services be hosted, and they cannot match the raw throughput or latency of direct socket connection~\cite{li2019socksdirect}.
\end{itemize}

The aforementioned advantages and disadvantages lead us to select the shared file system and in-memory data store (Redis) approaches to support intra-endpoint data transfers in \name{}, as these approaches are both general and are readily available (or can be deployed with minimal effort) on most target resources. 
We present a preliminary performance study of these four approaches in \S\ref{sec:data-eval} and the results show that the performance of shared file system and Redis is similar to the other approaches, especially when transferring large volumes of data.
We have extended the \name{} agent such that users may specify a requirement for a Redis cluster to be deployed alongside their endpoint. The \name{} SDK provides a general
interface to retrieve the Redis client which users can interact with, as shown in Listing~\ref{lst:redis}.

\begin{lstlisting}[style=PythonStyle, caption={Intra-endpoint data transfer with Redis. \label{lst:redis}}]
def example(key, data):
  from funcx_endpoint import get_redis_client
  rc = get_redis_client()
  rc.set(key, data)
  rc.get(key)
\end{lstlisting}

\section{Container Management}\label{sec:routing}
\name{} uses containers to provide customized execution environments
for functions irrespective of the endpoint's host environment. 
In this section, we discuss how the \name{} 
agent spawns containers to serve functions, retains warm containers, routes functions to containers, and scales resources based on function requirements.

\subsection{Container Warming}\label{sec:warm}
Commercial FaaS platforms~\cite{wang2018peeking}
keep function containers \emph{warm} by leaving them running for a short period of time (e.g., 5-15 minutes)
following the execution of a function.
Warm containers remove the need to instantiate a new container to execute a function, 
significantly reducing latency. 

We argue that this need is especially important in HPC environments for several reasons.
First, containers and Python environments (e.g., conda) are generally stored on shared file systems of HPC systems.
Therefore, starting many containers and Python environments concurrently for the workers at the HPC scale may impose significant stress on the shared file systems. Second, many HPC centers implement their own methods for instantiating containers that place
limitations on the number of concurrent requests.
Third, individual cores are often 
slower in many-core architectures like Xeon Phis. 
As a result, the start time for containers
can be much larger than what would be seen on a PC, as shown in \tablename~\ref{extractor-tab} in \S\ref{sec:eval-routing}.

In \name{}, container warming is implemented by the \name{} agent. To reduce the number of container cold starts, the \name{} agent keeps a container warm until there are insufficient resources available to process pending workloads or the container has been idle for a configurable period of time (e.g., 10 minutes).
The \name{} agent is extensible to support other container-warming strategies, such as releasing the least recently used container and application-agnostic container warming~\cite{shahrad2020serverless} if necessary. 

\subsection{Warming-aware Function Routing}

Ideally we aim to minimize the number of container cold starts due to the
cost of starting a container in HPC environments.
To do so, the \name{} agent needs to know which computing nodes have warm containers and what types of warm containers, so that it can route the function tasks to the appropriate warm containers.

The \name{} agent employs a hierarchical, warming-aware scheduling algorithm to route function tasks to workers to optimize throughput. The \name{} agent determines which functions to route to a given manager, and each \emph{manager} determines how to launch and spawn containers to satisfy the arriving workload.
Thus, warming-aware routing involves coordination between managers and \name{} agent.
Each manager advertises its deployed container types and its available resources to the \name{} agent. 
Based on the advertised information of each manager,
the \name{} agent implements a warming-aware scheduling algorithm to route tasks
to managers. Specifically, when receiving a task with requirement for a specific container type, the scheduler attempts to send the task to a manager that has a suitable warm container. 
When there are multiple available managers with the required container type warm, priority is given to the one with the most available container workers to balance load across managers.
If there are not any warmed containers on any connected managers, the \name{} agent chooses one manager at random to execute the task.
While we use random scheduling in our implementation, other scheduling policies, such as bin-packing and round-robin, could also be used.
To amortize  network latency during manager advertising and task dispatching, the \name{} agent also supports \emph{prefetching}, which allows a manager to prefetch a configurable number of additional tasks beyond its current availability. 

Upon receiving a set of tasks, the manager determines the required container types and dynamically starts (and stops) containers to serve tasks
in a fair manner: we set the number of deployed containers for a function type to be proportional to the number of received tasks of this function type. For instance, if 30\% of the tasks a manager receives are of type A and the manager can spawn at most 10 containers, the manager will spawn 3 containers of type A.

It is worth mentioning that the function routing is different when an endpoint is deployed on a Kubernetes cluster. Both the manager and its workers are deployed within a container pod that can only serve one type of function. Hence, in this case, each manager is deployed with a
specific container image and the agent simply needs to route
tasks to corresponding managers.

We apply relatively simple scheduling algorithms here to 
demonstrate the benefits of warming aware routing; however, the \name{} agent implements modular scheduling interfaces for function routing (at \name{} agents) and container deployment (by managers) which enabling
different algorithms (e.g., priority-aware or deadline-driven scheduling)
to be implemented by users.
We note that when task duration is much larger than the container cold start time, the benefits of warming-aware routing are limited.

\subsection{Elastic Resource Provisioning}

One of the main benefits of the FaaS computing model is elasticity.
To provide elasticity on a \name{} endpoint, a \name{} agent dynamically provisions 
resources via an extensible provisioning \emph{strategy} interface.

The strategy interface consists of a monitoring and a scaling component within the \name{} agent. The monitoring component interacts periodically (e.g., every second) with the provider interface (introduced in~\S\ref{sec:provider}) and the \name{} agent to fetch the current endpoint load, including the active and idle resources (i.e., the number of container workers) and the number of pending function requests.
Based on the monitoring information, the scaling component automatically provisions more resources when the number of function requests is greater than the number of idle resources, and releases resources that have been idle for some period of time, via the provider interface. The maximum idle time is set to two minutes by default, but is user-configurable for each endpoint.

Similar to commercial FaaS platforms such as AWS Lambda and Azure Functions, the \name{} strategy allows users to configure the minimum and maximum resources to be used, as well as how aggressively a \name{} agent scales those resources (e.g., request one more resource when there are ten waiting requests).
However, elasticity may be subject to resource request delays, such as the time to request a new instance on a cloud or 
to provision a resource via an HPC scheduler.

\section{Evaluation}\label{sec:evaluation}

We evaluate the performance of \name{} in terms of latency, scalability, and throughput. We also study the effects of batching, function routing, and data transfer approaches.

\subsection{Latency}

\begin{figure}[h]
  \includegraphics[width=0.97\columnwidth]{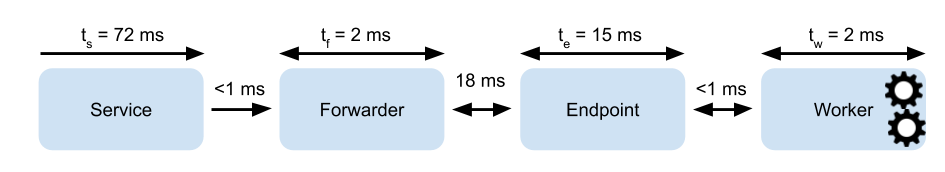}
    \vspace{-0.1in}
  \caption{\name{} latency breakdown for a container.}
  \label{fig:lat_breakdown}
    \vspace{-0.1in}
\end{figure}

We explore \name{} latency by instrumenting the system.
Figure~\ref{fig:lat_breakdown} shows latencies for a warm
container as follows:
$t_s$: Web service latency to authenticate, store the task in Redis, 
and append the task to an endpoint's queue;
$t_f$: forwarder latency to read task from the Redis store, forward 
the task to an endpoint, and write the result to the Redis store; 
$t_e$: endpoint latency to receive tasks and send results to the forwarder,
and to send tasks and receive results from the worker; and
$t_w$: function execution time.  
The endpoint was deployed on ANL's Cooley cluster for this test and had an 18 ms latency on average to the forwarder.
We observe that $t_w$ is fast relative to the overall system latency. 
The network latency between service and forwarder includes minimal 
communication time due to internal AWS networks (measured at $<$1 ms).
Most \name{} overhead is found
in $t_s$ due to  
authentication, and in $t_e$ due to internal queuing and 
dispatching.
We note here that the aim of \name{} is not to build yet another low-latency FaaS platform, but instead to provide a new federated model in which functions can be executed on arbitrary remote machines. Nevertheless, in our previous work we showed that the latency of \name{} is comparable to commercial FaaS platforms, such as AWS Lambda, Google Cloud Functions, and Azure Functions~\cite{chard2020funcx}. 

\begin{figure*}[h]
  \vspace{-0in}
  \centering
  \includegraphics[width=0.8\textwidth,trim=0.07in 0.08in 0.07in 0.08in,clip]{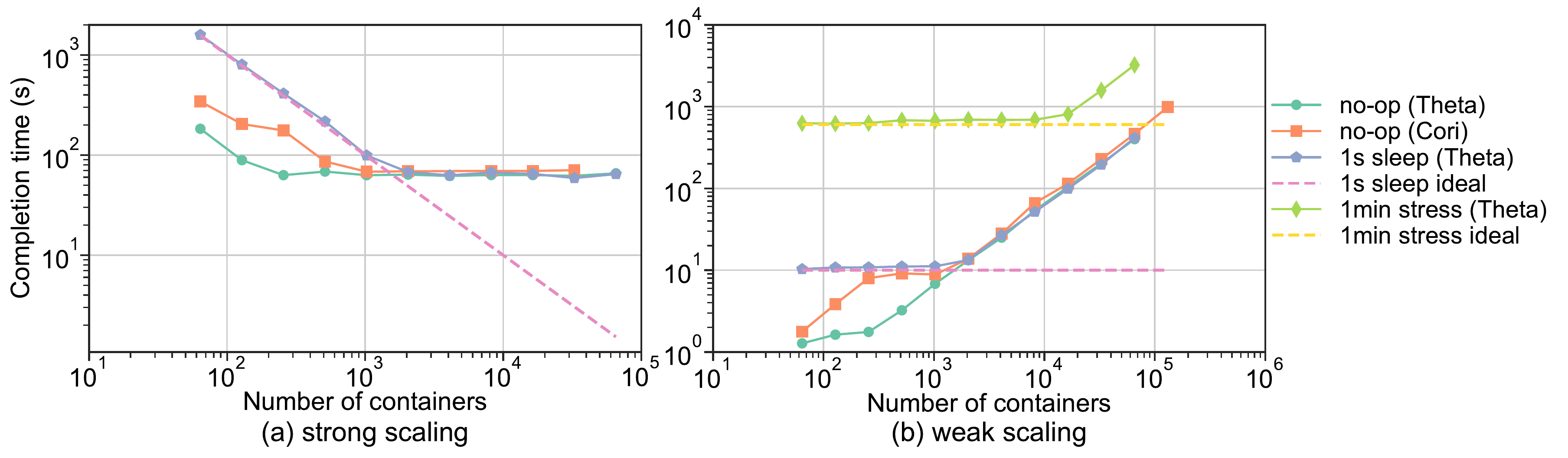}
  \vspace{-0.1in}
  \caption{Strong and weak scaling of the \name{} agent.}
\label{fig:scalability}
  \vspace{-0.1in}
\end{figure*}

\subsection{Scalability and Throughput}
We study the strong and weak scaling of the \name{} agent on  
ANL's Theta and NERSC's Cori supercomputers. 
Theta is a 11.69-petaflop system based on the second-generation 
Intel Xeon Phi ``Knights Landing" (KNL) processor. 
Its \num{4392} nodes each have a 64-core processor
with 16 GB MCDRAM, 192 GB of DDR4 RAM, and are interconnected with high speed InfiniBand.
Cori is a 30-petaflop system with an Intel Xeon ``Haswell" partition 
and an Intel Xeon Phi KNL partition. 
We ran our tests on the KNL partition, which 
has \num{9688} nodes, 
each with a 68-core processor (with 272 hardware threads) 
with six 16-GB DIMMs, 96 GB DDR4 RAM, 
interconnected in a 
Dragonfly topology.
We perform experiments using 64 Singularity containers on each Theta node 
and 256 Shifter containers on each Cori node.
Due to a limited allocation on Cori we use the four hardware threads
per core to deploy more containers than cores.

Strong scaling evaluates performance when the total number of function invocations is fixed; 
weak scaling evaluates performance when the average number of functions executed on each container is fixed. 
To measure scalability we created functions of various durations:  
a 0-second ``no-op'' function that exits immediately, a 1-second ``sleep'' function, and a 1-minute CPU ``stress'' function that keeps a CPU core at 100\% utilization.
For each case, we measured completion time of a batch of functions as we increased the total number of containers.
Notice that the completion time of running $M$ ``no-op'' functions on $N$ workers indicates the overhead of \name{} 
to distribute the $M$ functions to $N$ containers.
Due to limited allocations
we did not execute sleep or stress 
functions on Cori, nor did we 
execute stress functions for strong scaling on Theta.
We pre-warmed all containers in these experiments. 

\subsubsection{Strong scaling}

Figure~\ref{fig:scalability}(a) shows the completion time of \num{100000} 
\emph{concurrent} function requests with an increasing number of containers.
On both Theta and Cori, the completion time decreases as the number of containers 
increases, until we reach 256 containers for ``no-op'' and 2048 containers for ``sleep'' on Theta.
As reported by Wang et al.~\cite{wang2018peeking} and Microsoft~\cite{azureFunctionsDocs}, 
Amazon Lambda achieves good scalability for a single function to more
than 200 containers, Microsoft Azure Functions 
can scale up to 200 containers, and Google Cloud Functions does not scale well beyond 100 containers.
While these results do not indicate the maximum number of containers that can be used
for a single function, and likely include per-user limits imposed by the platform, 
our results show that \name{} scales similarly to commercial platforms.

\subsubsection{Weak scaling}
We performed \emph{concurrent} function 
requests such that each container receives, on average, 10 requests.
Figure~\ref{fig:scalability}(b) shows weak scaling for ``no-op,'' ``sleep,'' and ``stress.'' 
For ``no-op," the completion time increases with more containers on both Theta and Cori. 
This reflects the time required to distribute requests to all of the containers.
On Cori, \name{} scales to \num{131072} concurrent containers and 
executes more than 1.3 million ``no-op'' functions.
Again, we see that the completion time for ``sleep'' remains close to constant 
up to 2048 containers, and the completion time for ``stress'' remains 
close to constant up to \num{16384} containers. 
Thus, we expect a function with a several-minute duration would scale well to many more containers. 

\subsubsection{Throughput}
We observe a maximum throughput for a \name{} agent (computed as number of function requests divided 
by completion time) of \num{1694} and \num{1466} requests per second on Theta and Cori, respectively.

\subsubsection{Summary}
Our results show that \name{} agents 
i) scale to \num{130000}+ containers for a single function; 
ii) exhibit good scaling performance up to approximately \num{2048} containers for a 
1-second function and \num{16384} containers for a 1-minute function; 
and iii) provide similar scalability and throughput using both Singularity and 
Shifter containers on Theta and Cori. It is important to note that these experiments study the
\name{} agent, and not the end-to-end throughput of \name{}. 
While the \name{} web service can elastically scale
to meet demand, 
communication overhead may limit throughput. To address this challenge and 
amortize communication overheads, we enable batch submission of tasks. 
These optimizations are discussed in \S\ref{sec:opt}.

\subsection{Data Management}\label{sec:data-eval}
\begin{figure}[h]
  \vspace{-0in}
  \includegraphics[width=0.9\columnwidth,trim=0.07in 0.08in 0.07in 0.08in,clip]{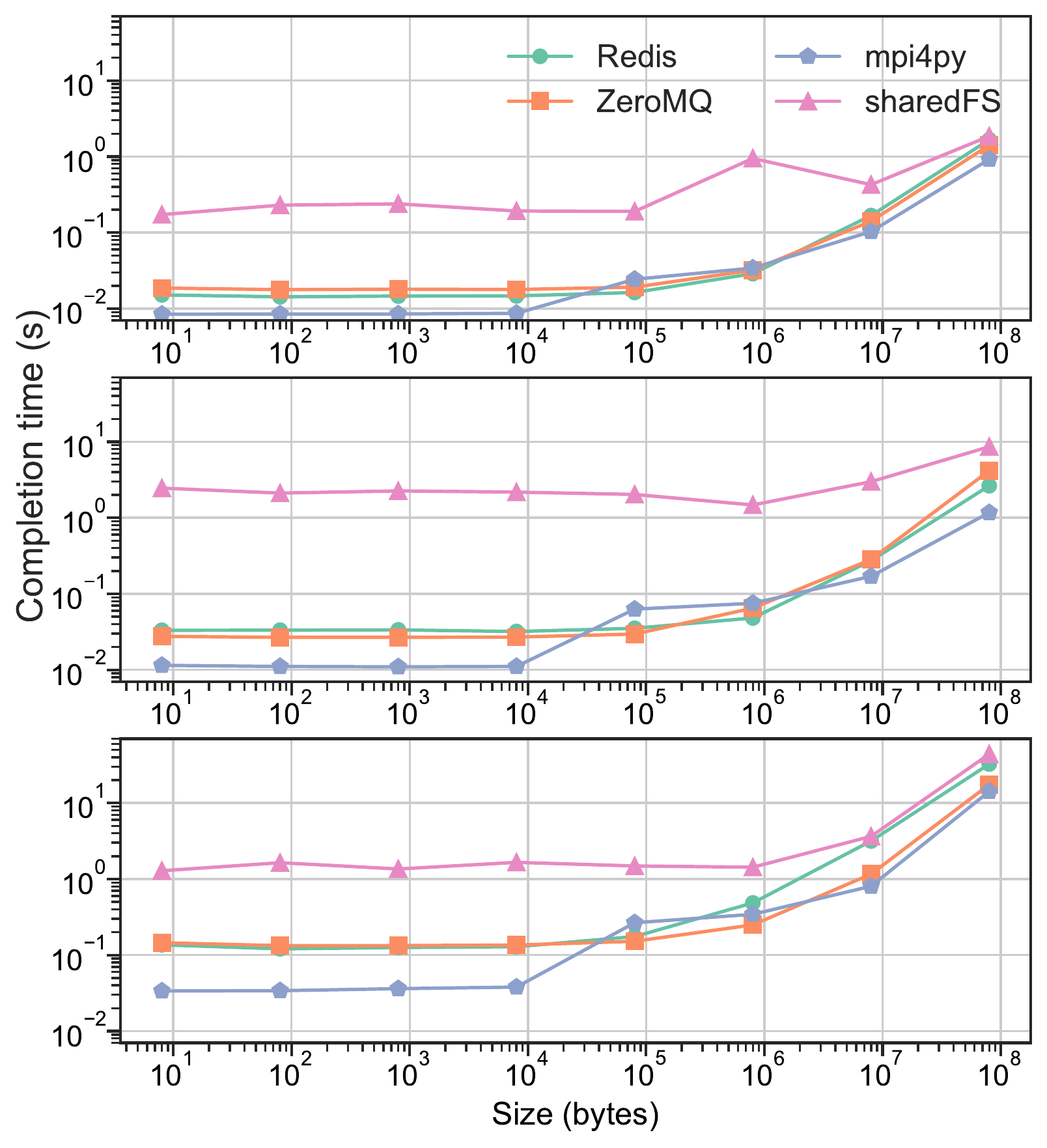}
  \vspace{-0.1in}
  \caption{Performance of the four intra-endpoint transfer approaches. Top: point-to-point; middle: broadcast to 20 nodes; bottom: all-to-all on 20 nodes.}
\label{fig:redis-broadcast}
  \vspace{-0.1in}
\end{figure}

We evaluate four potential approaches for intra-endpoint data transfers (described in \S\ref{sec:intra}) on  Theta. We use mpi4py for MPI, ZeroMQ for direct socket connections, Redis for the in-memory store, and Theta's Lustre shared file system. We note that we use mpi4py because it supports direct Python object transfers and previous work~\cite{saxena2018evaluation} has shown that mpi4py does not add significant overhead when compared with OpenMPI in terms of throughput and latency for data transfers.
We emulate different communication patterns (i.e., point-to-point, broadcast, and all-to-all) and vary data transfer size.
\figurename~\ref{fig:redis-broadcast} shows the performance of these four approaches with different communication patterns. As expected, MPI performs the best, and sharedFS the worst. 
However, ZeroMQ and Redis achieve similar performance to MPI. As 
data volume increases, the performance difference between the four approaches diminishes, as transfer time is mainly determined by available network bandwidth (which is the same for all approaches). 

While sharedFS and Redis perform slightly worse than MPI for small data sizes, we adopt them in \name{} because of their generality, ease-of-use, and the challenges of using mpi4py (as well as MPI compiled in C) and ZeroMQ described in \S\ref{sec:intra}.

We now evaluate intra-endpoint data management in the context
of MapReduce applications and a real-world science application 
deployed on funcX.

\subsubsection{MapReduce}
To demonstrate how Redis and sharedFS can facilitate intra-endpoint data transfers for real applications, we deployed a \name{} endpoint with a three-node Redis cluster. We also used the shared Lustre file system on Theta. We deployed two MapReduce applications: WordCount and Sort. These applications involve an all-to-all communication pattern between map and reduce tasks (i.e., data shuffling). 

Each application processes 30~GB of Wikipedia text data, and has 300 map and 300 reduce tasks, requiring communication of 90000 data chunks in total.
\tablename~\ref{tab:mr-redis} shows the average completion time of each task spent in each phase of the MapReduce application: input read, map process, intermediate write, intermediate read, reduce process, output write, when using Redis and sharedFS approaches for data shuffling. WordCount benefits less from Redis than Sort as WordCount shuffles just one tenth of the data. The table shows that Redis can speed up the data shuffling phase of the workload (i.e., intermediate write and read) by up to 3x.

\begin{table}[h]
\caption{Average completion time of the transfer phases in WordCount and Sort when using Redis and shared file system for intra-endpoint data management.}
\label{tab:mr-redis}
\begin{tabular}{lll|ll}
 & \multicolumn{2}{c|}{\textbf{WordCount (s)}} & \multicolumn{2}{c}{\textbf{Sort (s)}} \\
\multicolumn{1}{c}{} & \multicolumn{1}{c}{\textbf{Redis}} & \multicolumn{1}{c|}{\textbf{SharedFS}} & \multicolumn{1}{c}{\textbf{Redis}} & \multicolumn{1}{c}{\textbf{SharedFS}} \\ \hline
Intermediate write & 3.55 & 8.15 & 3.27 & 5.32 \\
Intermediate read & 33.39 & 43.40 & 11.37 & 41.77 \\
\end{tabular}
\end{table}

Note that \tablename~\ref{tab:mr-redis} shows the average task completion time.
The benefits of Redis over sharedFS on the total completion time of a MapReduce application may depend on the amount of parallelism and the portion of the shuffle phase over the other phases.
For example, the total completion time of Sort with Redis and sharedFS are 
220 and 520 seconds, respectively, yielding a 55.7\% improvement. The WordCount application runs in 1800 seconds and 2200 seconds, respectively, yielding a 18.2\% improvement. This is because Sort has a heavier shuffle phase than WordCount.

\subsubsection{Colmena}

Finally, we evaluate intra-endpoint data management in the context of a  real-world scientific application to demonstrate the benefits of Redis over sharedFS. Colmena~\cite{ward2021colmena} is a framework that manages large-scale, AI-directed steering of computational campaigns (e.g., to efficiently explore large molecular spaces when designing new materials). A Colmena application consists of a \emph{Thinker} that implements the decision-making policy to generate new tasks (e.g., new simulation, new model training, or model inference), a \emph{Task Server} that dispatches task requests to resources and manages task results, and \emph{Workers} that are deployed on compute resources to execute tasks. 
These components exchange data (e.g., task requests and results) 
with Redis used to facilitate transfers. 
We implement a Colmena benchmark with 1000 tasks, each with 1 MB input and 1 MB output data. Table~\ref{tab:colmena} shows the average completion time of the communication stages in Colmena. 
Redis yields a lower completion time for all communication stages compared to sharedFS. Such a benefit has been demonstrated to be particularly important when running Colmena at scale with thousands of tasks.

\begin{table}[h]
\centering
\caption{Average completion time of the transfer phases in the Colmena benchmark when using Redis and shared file system for intra-endpoint data management.}
\label{tab:colmena}
\begin{tabular}{c c c}
 \textbf{Stage} & \textbf{Redis (ms)} & \textbf{SharedFS (ms)}\\ \hline
Input data write from Thinker & 7.15 &  32.31 \\ 
Input read on Workers & 0.70 &  11.36 \\ 
Result write from Workers & 18.04 & 244.72 \\ 
Result read from Task Server & 0.11 & 3.50 \\ 
\end{tabular}
\end{table}

\begin{figure*}[h]
  \includegraphics[width=0.97\textwidth]{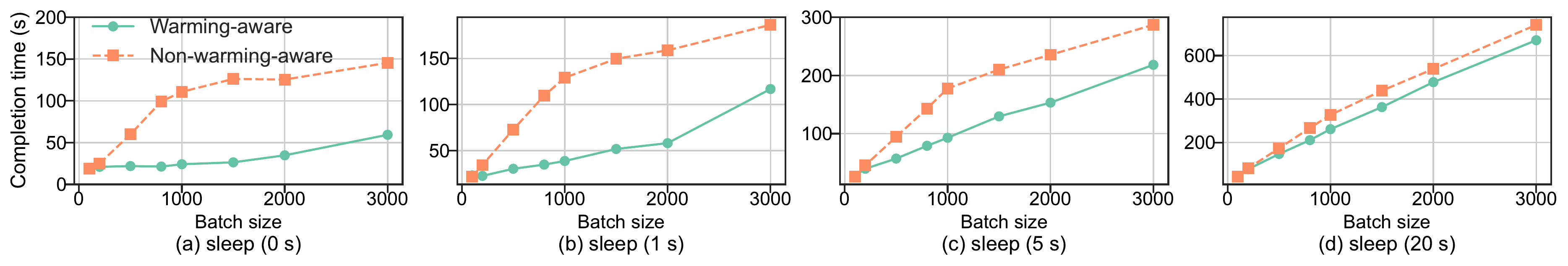}
    \vspace{-0.1in}
  \caption{Completion time of warming-aware and non-warming-aware routing.}
  \label{fig:routing-throughput}
    \vspace{-0.2in}
\end{figure*}

\begin{figure*}[h]
  \includegraphics[width=0.97\textwidth]{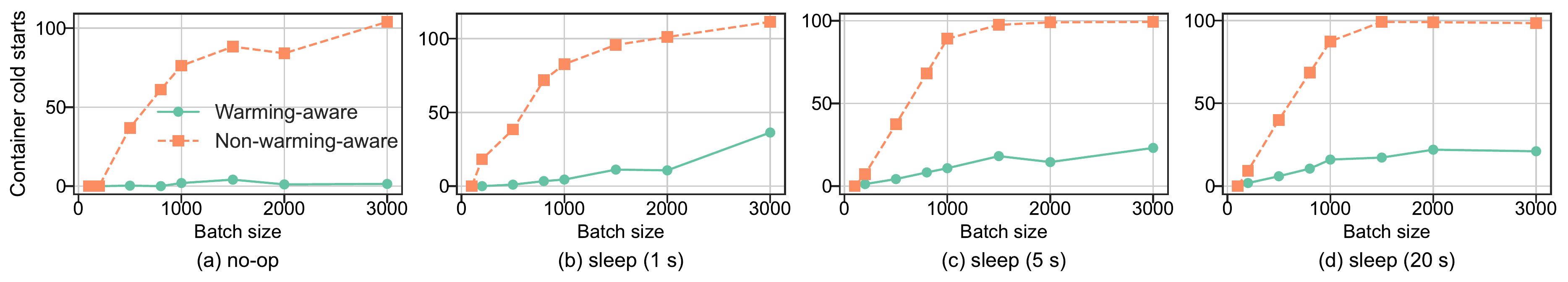}
    \vspace{-0.1in}
  \caption{Number of container cold starts of warming-aware and non-warming-aware routing.}
  \label{fig:routing-switch}
    \vspace{-0.1in}
\end{figure*}

\subsection{Function Routing}\label{sec:eval-routing}
Before exploring function routing performance, we first quantify the instantiation cost of various container technologies on different resources. 
Specifically, we measure the time taken to start a container and execute a Python command to import \name{}'s worker module---a
requirement prior to executing a \name{} function.
We deploy the containers on an AWS EC2 \texttt{m5.large} instance and on compute nodes on Theta and Cori following the facility's documented
best practices.
\tablename~\ref{table:cold_start_cost}
shows the results.
We speculate that the significant performance deterioration of container instantiation on HPC systems can be attributed
to a combination of slower clock speed on KNL nodes and shared file system contention when fetching images.
These results highlight the need to apply function warming approaches to reduce overheads on HPC systems.

\begin{table}[h]
\caption{Cold container instantiation time for different container technologies on different resources.}
\label{extractor-tab}
 \vspace{-0.1in}
\begin{center}
  \begin{tabular}{l l c c c}
    \textbf{System} & \textbf{Container} & \textbf{Min (s)} & \textbf{Max (s)} & \textbf{Mean (s)} \\
    \hline
    Theta & Singularity & 9.83 & 14.06 & 10.40  \\
    Cori & Shifter & 7.25     & 31.26    & 8.49 \\
    EC2 & Docker   & 1.74     & 1.88     & 1.79 \\
    EC2 & Singularity & 1.19  & 1.26     & 1.22 \\
  \end{tabular}
\end{center}
  \vspace{-0.1in}
\label{table:cold_start_cost}
\end{table}

We evaluate \name{}'s function routing strategy and show that it improves
overall throughput as well as reducing the number of container cold starts.
We deployed an endpoint on Theta and compared the performance of warming-aware routing and randomized (non-warming-aware) routing.
The endpoint is allocated 10 nodes and each node can host 10 workers, each with its own container. 
We registered 10 functions, where each function requires a specific container (i.e., 10 different containers.) 
We submitted a batch of requests, each of which is chosen from one of the ten functions uniformly at random.
\figurename~\ref{fig:routing-throughput} and \figurename~\ref{fig:routing-switch} show the overall function completion time and the number of container cold starts for different batch sizes and for different function durations (0, 1, 5, and 20 seconds). 
We note that the number of requests in a batch is much higher than the available resources (100 container workers) in this experiment, and thus a container worker is more likely to be killed to serve other request when using non-warming-aware routing.
Thus, the warming-aware routing reduces completion time by up to 61\% for a batch of requests (i.e., higher throughput) and reduces the number of container cold starts significantly (e.g., 22 cold starts for 3000 functions), compared to the randomized routing strategy. This is because the warming-aware algorithm attempts to reuse the warm containers as much as possible to reduce the overhead of container instantiation. 
As expected, the benefit of warming-aware routing gradually diminishes as the function duration increases, because the function runtime, rather than the cold container instantiation time, becomes dominant.

\subsection{Batching}
\label{sec:opt}
To evaluate the effect of executor-side batching, we submit \num{10000} concurrent ``no-op'' function requests
and measure the completion time when executors can request one function at a time (batching disabled) vs
when they can request many functions at a time based on the number of idle containers (batching enabled). 
We use 4 nodes (64 containers each) on Theta.
We observe that the completion time with batching enabled is 6.7~s (compared to 118s when disabled).

\section{Experiences with \name{}} \label{sec:discussion}

As of August, 2022 \name{} has been used by 413 users to perform over 19.8 million function invocations, \num{338105} functions have been registered, and 4027 endpoints have been created.
Here we describe our experiences applying \name{} to various scientific case studies.

\textbf{AI-enabled steering of computational campaigns:} 
Colmena~\cite{ward2021colmena} is an open-source library that enables researchers to build complex, AI-directed HPC campaigns.
Researchers can implement flexible decision-making policies to steer different tasks (e.g., simulation, model update, and model inferences) of computational campaigns. When tasks are generated, \name{} serves as an execution backend to distribute and execute tasks. The FaaS model of \name{} and the implementation of container management allows Colmena to flexibly dispatch tasks to arbitrary computing resources, enabling ML-enhanced tasks to be sent to GPU-accelerated devices and high throughput simulations to HPC clusters. The integration of data management mechanisms (e.g., Globus and Redis) in \name{} enables data to move between Colmena entities transparently without requiring the user to manage movement; further, it can improve performance and simplify distributed, data-intensive campaigns.

\textbf{Linking instruments and HPC:}
\name{} has been used to combine several experimental instruments with HPC infrastructure~\cite{vescovi2022linking}. 
This approach allows scientists to offload computationally-intensive analysis tasks to HPC resources, simplifies large-scale parallel processing for large data rates, and enables online analysis. Such experimental techniques, including serial synchrotron crystallography~\cite{sherrellfixed2022}, X-ray photon correlation spectroscopy~\cite{perakis2020towards}, ptychography~\cite{bicer2021high}, and scientific machine learning~\cite{liu2022braggnn},  
depend on orchestration of various activities in various locations. For this purpose, these examples use Globus flows~\cite{chard2022globus} to create complex sequences of actions. For example, when data are acquired from an experiment, run quality control at the edge, move data to an HPC center, run analysis and reconstruction algorithms, and index resulting images in a data catalog. \name{} provides the compute substrate enabling many of these actions to be executed in various locations. Integration of Globus for data management simplifies dispatching tasks to different resources without requiring changes to broader workflows to transfer and retrieve inputs and results. Further, scientific analysis toolkits are often containerized to promote portability and exploit available resources. 
The container warming features presented in this paper 
enable these workloads to reduce cold starts, which can be costly on large, shared file systems and facilitate rapid computation---a necessary feature to support real-time computation and experiment steering. We report function execution and data transfer characteristics in prior work~\cite{vescovi2022linking}.

\textbf{AlphaFold as a Service:}
AlphaFold~\cite{jumper2021highly} is a cutting edge machine learning model that can predict a protein's 3D structure from its amino acid sequence. AlphaFold has garnered significant interest in the bioinformatics community, with applications in the development of therapeutics and accelerating the practice of deriving crystal structures at light sources. However, AlphaFold relies on powerful GPUs and 
large reference datasets, restricting access for many researchers. To address these challenges, the Argonne Leadership Computing Facility deployed AlphaFold as a service using \name{} to dynamically provision GPU resources on-demand. In this work, containers are dispatched to GPU nodes and managed by the \name{} endpoint to serve inference requests. As AlphaFold tasks can take over an hour to process, the Globus integration with \name{} provides the ability to asynchronously transfer results to users.

\textbf{Distributed ML:}
Flox~\cite{kotsehub22flox} is a federated learning (FL) framework that decouples model training and inference from infrastructure management. 
Flox uses \name{} to enable users to 
train and deploy FL models on one or more remote computers, and in particular on edge devices. We are applying these techniques to Rural AI applications~\cite{ruralai2022}, using \name{} to 
facilitate training and deployment of models in remote locations. 
Rural AI requires reliable task and result transmission
as devices are deployed in
rural settings where device and network outages are common, 
and the quality of wireless networks varies depending on location. 
\name{}'s hierarchically
designed queues support the necessary robustness to dispatch (and queue) tasks across rural devices. Containers enable execution of tasks on heterogeneous
edge devices for training and on centralized cloud instances for model
aggregation.
\section{Related Work}\label{sec:survey}
Both commercial and open-source FaaS platforms have proved extremely successful in industry as a way
to reduce costs and remove the need to manage infrastructure.

\textbf{Hosted FaaS platforms:}
\emph{Amazon Lambda}~\cite{amazonlambda},
\emph{Google Cloud Functions}~\cite{googlecloudfunctions}, and \emph{Azure Functions}~\cite{azureFunctions} are the most well-known FaaS platforms. They support various function languages
and trigger sources, connect directly to other cloud services, and apply fine-grain billing models.  Lambda uses Firecracker~\cite{agache2020firecracker}, a custom virtualization technology built on KVM, to create lightweight micro-virtual machines. 
To meet the needs of IoT use cases,
some cloud-hosted platforms support local deployment (e.g., AWS Greengrass~\cite{greengrass}); 
however, they support only single machines and require that functions be exported from the cloud platform.

\textbf{Open source platforms:}
Open FaaS platforms resolve two of the key challenges to using FaaS for scientific workloads: they 
can be deployed on-premise and can be customized to meet the requirements of data-intensive 
workloads without set pricing models.

\emph{Apache OpenWhisk}~\cite{openwhisk}, the basis of IBM Cloud Functions~\cite{IBMCloudFunctions},
defines an event-based programming model, consisting of \emph{Actions} which are stateless, runnable functions, \emph{Triggers} which are the types of events OpenWhisk may track, and \emph{Rules} which associate one trigger with one action. OpenWhisk can be deployed locally as a service using a Kubernetes cluster. 

\emph{Fn}~\cite{Fn} is an event-driven FaaS system that executes functions in Docker containers. Fn allows users to logically group functions into applications. 
Fn can be deployed locally (on Windows, MacOS, or Linux) or on Kubernetes.

The \emph{Kubeless}~\cite{Kubeless} FaaS platform builds upon Kubernetes. 
It uses Apache Kafka for messaging, provides a CLI that mirrors Amazon Lambda, and supports comprehensive monitoring. 
Like Fn, Kubeless allows users to define function groups that share resources.

\emph{SAND}~\cite{akkus2018sand}, which has been recently open-sourced as KNIX MicroFunctions~\cite{knix}, is a lightweight, low-latency FaaS platform from Nokia Bell Labs that provides application-level sandboxing and a light-weight process-based execution model. KNIX provides support for function chaining via user-submitted workflows. 
Recently, KNIX has been further extended to support GPU sharing among functions~\cite{satzke2021efficient}. 
However, KNIX requires privileged access to nodes, which is generally  not possible in research CI.

\emph{Abaco}~\cite{stubbs2017containers} implements the Actor model, where an \textit{actor} is an Abaco runtime mapped to a specific Docker image. Each actor executes in response to messages posted to its \textit{inbox}. It supports functions written in several programming languages and automatic scaling. Abaco also provides fine-grained monitoring of container, state, and execution events and statistics. Abaco is deployable via Docker Compose.

\emph{ChainFaaS}~\cite{ghaemi2020chainfaas} is a blockchain-based FaaS platform that makes use of idle personal computers. The platform allows users to submit functions that utilize contributed computing power, or to be a provider who contributes the idle computing resources for potential profits. While ChainFaaS shares some similar goals with funcX, it focuses on deployment on personal computers, rather than large-scale research CI.

\emph{DFaaS}~\cite{ciavotta2021dfaas} is a federated and decentralized FaaS platform for edge computing. It relies on a peer-to-peer network to share the states of edge nodes to balance loads among all the nodes.

\textbf{Comparison with \name{}:}
Hosted cloud providers implement high performance and reliable FaaS models 
that are used by an enormous number of users. However, they often have vendor lock-in, are not designed
to support heterogeneous resources or research CI (e.g., schedulers, containers), 
do not integrate with the science ecosystem 
(e.g., in terms of data and authentication models), and can be costly. 

Open source and academic frameworks support on-premise deployments and can be 
configured to address a range of use cases. However, most systems we surveyed
are Docker-based and rely on Kubernetes (or other
container orchestration platforms) for deployment.
Some systems such as ChainFaaS and DFaaS support distributed function execution on personal computers and edge nodes. 
However, to the best of our knowledge, there are no systems that support remote execution over a federated ecosystem of endpoints on diverse research CI (from edge to HPC environments).

\textbf{Other Related Approaches:}
FaaS has many predecessors, notably
grid and cloud computing, container orchestration, and analysis systems. 
Grid computing~\cite{Foster2001} laid the foundation for 
remote, federated computations, often through
federated batch submission~\cite{krauter02gridmanagement}. 
GridRPC~\cite{seymour02gridrpc} defines an API for executing 
functions on remote servers requiring that developers implement
the client and the server code. 
\name{} extends these ideas to allow interpreted functions
to be registered and then executed within sandboxed 
containers via standard cloud and endpoint APIs.

Container orchestration systems, such as Mesos~\cite{hindman11mesos}, Kubernetes~\cite{hightower17kubernetes}, KubeFed~\cite{kubefed}, MicroK8s~\cite{microk8s}, and K3s~\cite{k3s}, allow
users to scale deployment of containers while 
managing scheduling, fault tolerance, resource provisioning, and addressing
other user requirements. 
Mesos and Kubernetes primarily rely on dedicated, cloud-native
infrastructure. KubeFed extends Kubernetes to support multi-cluster deployments.
MicroK8s and K3s are lightweight versions of Kubernetes and are designed for Edge and IoT use cases.
These systems cannot be directly used with diverse research CI (e.g., HPC resources); however, these container orchestration systems serve as a basis for developing serverless platforms, such as Kubeless, and indeed play an increasingly important
role in research CI.
\name{} focuses at the level of scheduling and managing functions, 
that are deployed across a pool of containers.
We leverage both container orchestration
systems (e.g., Kubernetes) as well as techniques from 
orchestration systems (e.g., warming) in funcX.

Data-parallel systems such as Hadoop~\cite{hadoop} and Spark~\cite{spark}
enable map-reduce style analyses. 
Unlike \name{}, these systems dictate a particular programming
model on dedicated clusters. 
Python parallel computing libraries such as Parsl and Dask~\cite{dask} 
support development of parallel programs, 
and parallel execution of selected functions within those scripts, 
on clusters and clouds. These systems could be extended to use \name{}
for remote execution of tasks.

LFM~\cite{shaffer2021lfm} provides advanced dependency management for Python functions by using transparent dependency detection and distribution, and dynamic provisioning and resource management at the granularity of a Python function. 
Azure Functions~\cite{shahrad2020serverless} proposed a policy that dynamically controls the pre-warming window for application containers to reduce the number of container cold starts, based on the characterization of applications.
Researchers have proposed various methods to mitigate container cold start latency by leveraging various workflow-specific information, such as cascading starts and dependency graphs~\cite{daw2020xanadu,shen2021defuse,bermbach2020using}.
Anna~\cite{sreekanti2020cloudburst} is an autoscaling key-value store that can be used to support stateful serverless computing.
Delta~\cite{kumar2021delta} adds a shim layer on top of \name{} that profiles the function performance on different endpoints and automatically schedules functions to appropriate endpoints.
Several recent papers have aimed to model application performance and optimize performance on FaaS platforms~\cite{HoseinyFarahabady2018model,bao2019performance,kim2020automated,lin2021modeling}.
While \name{} implements its own function routing, container management, data management schemes, and performance metrics, 
these systems are orthogonal to this paper and could be integrated with \name{}.

Several frameworks have been implemented on top of \name{} to create workflows for different scientific use cases. For instance, Xtract~\cite{skluzacek2021serverless} uses \name{} to enable workflow compositions for distributed bulk metadata extraction. Globus Automate~\cite{vescovi2022linking} uses \name{} to run arbitrary computations as part of automated and event-based workflows, it uses \name{}'s APIs to automatically monitor the status of a \name{} function and trigger the next step when it completes.

\section{Conclusion}\label{sec:conclusion}

\name{} is a distributed FaaS platform that is designed to support the unique
needs of research
computing. Unlike existing centralized FaaS platforms, 
\name{} combines a reliable and easy-to-use cloud-hosted
interface with the ability to securely execute functions on user-deployed \name{}
endpoints deployed on various remote computing resources. 
\name{} supports many HPC systems and cloud platforms, can use three container 
technologies, and can expose access to heterogeneous and specialized computing resources. 
In this paper we extend \name{} to support inter-endpoint and intra-endpoint data transfers between functions, and optimize function execution performance with advanced container management and warming-aware function routing mechanisms. 
We showed that \name{} agents can 
scale to execute 1M tasks over \num{130000} concurrent workers 
when deployed on the Cori supercomputer.  
We also showed that \name{}'s data transfer mechanisms are comparable to alternative methods, and that they can significantly improve application performance. 
Finally, we showed that \name{} can dynamically route functions to workers to reduce container warming overhead
and that batching can significantly reduce overheads. 

\name{} demonstrates the advantages of adapting the FaaS model to create
a federated computing ecosystem. Based on early experiences using \name{} in 
scientific case studies~\cite{chard2020funcx},
we have found that the approach
provides several advantages, 
including abstraction, code simplification, portability, scalability, and sharing; however, 
we also identified several limitations including suitability for some applications, 
conflict with current allocation models, and challenges decomposing
applications into functions. 
We hope that \name{} will serve as a flexible platform for research 
computing while also enabling new studies in 
function scheduling, 
dynamic container management, and data management.

In future work, 
we will continue our work to explore new scheduling approaches that can select appropriate endpoints for function execution and manage  data dependencies between functions. We also plan to provide APIs that allow users to manage and discover functions and endpoints.
We will extend \name{}'s container management capabilities to 
create containers dynamically based on function requirements,
and to stage containers to endpoints on-demand.
We will also explore techniques for optimizing performance, for example by
sharing containers among functions with similar dependencies and developing
resource-aware scheduling algorithms.

\section*{Acknowledgment}
This work was supported in part by NSF 2004894/2004932 and Laboratory Directed Research and
Development funding from Argonne National Laboratory under U.S. Department of
Energy under Contract DE-AC02-06CH11357. This work also used resources of the
Argonne Leadership Computing Facility and Center for Computational Science and Engineering at Southern University of Science and Technology.

\bibliographystyle{abbrv}
\bibliography{Bibs/funcx-refs}
\end{document}